\documentclass[preprint2,twoside]{hwo}

\usepackage{lipsum}

\input{hwo.h}

\begin{document}

\title{\textbf{\LARGE Unveiling supermassive black hole binaries with FUV-to-NIR spectropolarimetry}}
\author {\textbf{\large Frédéric Marin,$^{1}$ Julie Biedermann,$^1$ Thibault Barnouin,$^1$}}
\affil{$^1$\small\it Universit\'e de Strasbourg, CNRS, Observatoire Astronomique de Strasbourg, UMR 7550, 11 rue de l'universit\'e, 67000 Strasbourg, France}



\begin{abstract}
In this proceedings, we review the importance and complexity of detecting and characterizing supermassive binary black holes using conventional techniques (spectroscopy, timing, imaging). We show how spectropolarimetric data can strengthen or discard binary black hole candidates on solid grounds, and present what kind of instrument could perform this task. Namely, a high spectral resolution polarimeter operating from the far-ultraviolet (FUV) to the near-infrared (NIR), such as the POLLUX prototype, would perfectly fit within the payload of the Habitable Worlds Observatory and help revolutionize the study of extragalactic objects.
\\
\\
\\
\end{abstract}

\vspace{2cm}

\section{Science Goal: Unveiling galaxy evolution through the mergers of supermassive black holes}

Supermassive black holes (SMBHs) are the titanic engines that power the centers of galaxies, shaping their evolution and the large-scale structure of the Universe. When two galaxies collide and merge, their central black holes are expected to form a binary system before ultimately coalescing. These mergers are fundamental to our understanding of galaxy evolution, accretion physics, and even gravitational waves. Yet, detecting these binary SMBHs has proven extremely difficult.

Most detection methods rely on indirect signatures, such as periodic light variations or unusual spectral features, but these can often be explained by other astrophysical processes. Polarimetry, however, provides an independent and powerful way to reveal binary SMBHs. By measuring the polarization of light (how its electric and magnetic waves are oriented), we can detect unique signatures caused by the interaction of the two black holes' accretion disks and surrounding material. What makes this technique particularly powerful is that those features cannot be mimicked by a single black hole system.

Using a next-generation far-ultraviolet (FUV) to near-infrared (NIR) spectropolarimeter like POLLUX \citep{Neiner2025}, it will become possible to identify telltale polarization signatures of binary SMBHs, confirming their presence and characterizing their properties. This technique could revolutionize our ability to find these elusive systems, bridging the gap between electromagnetic observations and gravitational wave detections. Understanding binary SMBHs will not only shed light on black hole mergers but also help to decipher the cosmic history of galaxies and their monstrous central engines.

The fundamental question of “How SMBH mergers shape galaxy evolution” directly addresses key priorities laid out in the Astro2020 Decadal Survey. In particular, it supports the following Science Panel Questions: “What seeds supermassive black holes and how to they grow?”, and “How do supermassive black holes form, and how is their growth coupled to the evolution of their host galaxies?” since binary SMBHs are central to shaping galaxy evolution through both mergers and feedback. Furthermore, it aligns with the Science Panel Discovery Area “Transforming Our View of the Universe by Combining New Information from Light, Particles, and Gravitational Waves”, as spectropolarimetric observations of binary SMBHs offer early identification and environmental characterization of the systems that will later be detected in gravitational waves by, e.g., the Laser Interferometer Space Antenna (LISA, \citealt{Amaro2023}) and pulsar timing arrays \citep{Romani1989}. The capabilities of POLLUX aboard NASA’s future High-Definition Space Telescope will be uniquely suited to deliver on these goals, offering a powerful new tool to advance the decadal vision.

\section{Science Objectives}

\subsection{Identify distinct polarization signatures of binary SMBHs}

The fundamental goals of this science case are to identify SMBHs in the electromagnetic domain and to observe the asymmetries in their accretion structures. These are crucial for understanding how black holes grow, how they interact with their galactic environments, and how they contribute to galaxy evolution. Indeed, binary SMBHs are expected outcomes of galaxy mergers, a key driver of hierarchical structure formation \citep{Kauffmann2000}. However, their detection remains challenging due to their compact nature and the limited resolution of current instruments. Most existing methods rely on indirect evidence -- such as periodic flux variations or velocity offsets in spectral lines \citep{Zheng2023,Pasham2024} -- that can often be mimicked by single-SMBH systems or other astrophysical phenomena \citep{Vaughan2016,Zhu2020}. Polarimetry, on the other hand, offers a unique and independent pathway to identify these binaries and probe their physical structure, thanks to its sensitivity to geometry and scattering mechanisms.

\begin{figure*}[ht]
\includegraphics[width=\textwidth]{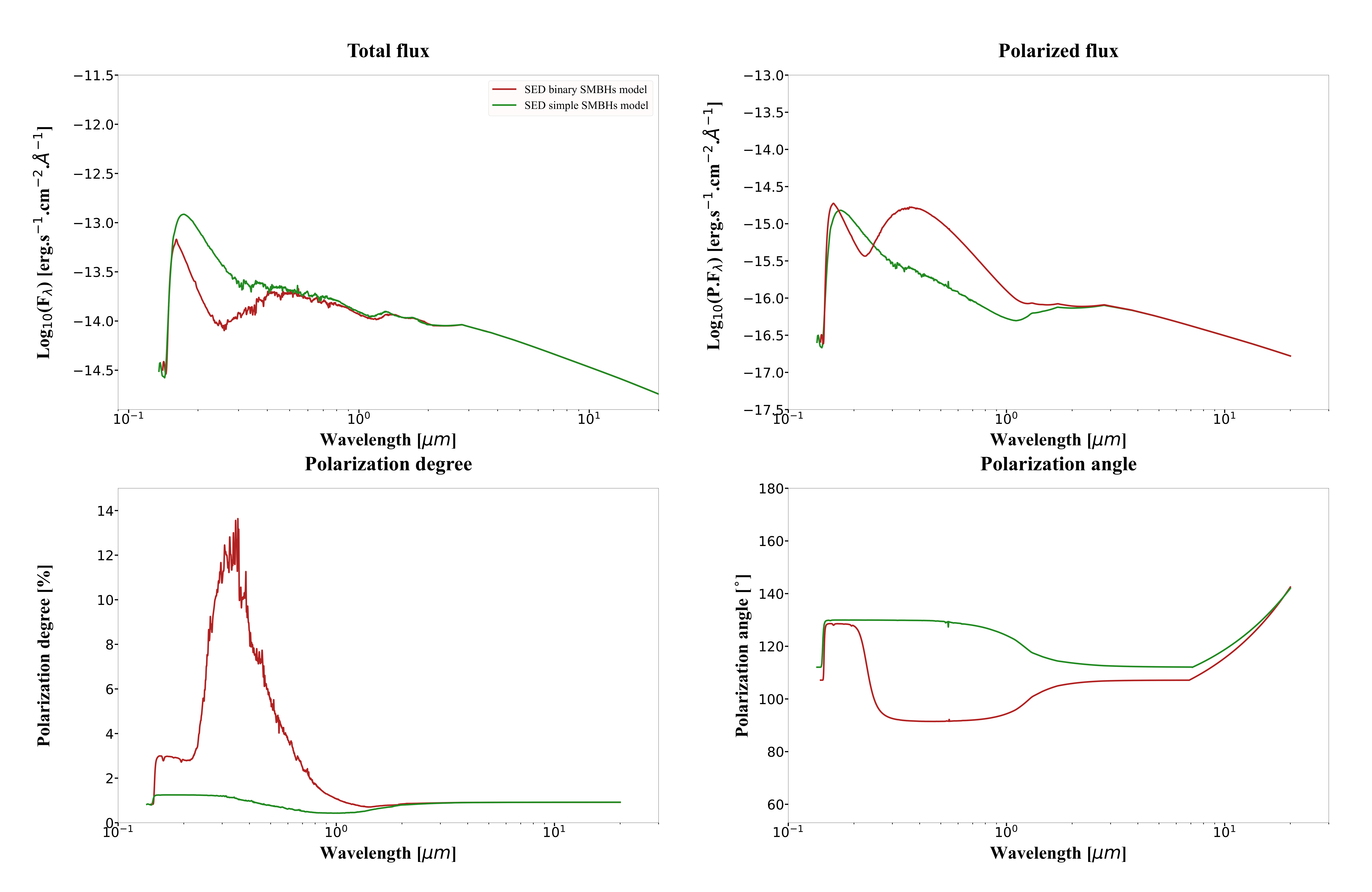}
\caption{\small Expected spectral energy distribution (SED, top left), polarized SED (top-right), polarization degree (bottom-left) and polarization angle (bottom-right) for an AGN seen from the top with a binary SMBHs system in its core (in red), compared to a standard pole-on AGN with a single SMBH (in green).
\label{author_fig1_label}
}
\end{figure*}

Indeed, binary SMBHs produce specific spectropolarimetric features due to their asymmetric accretion structures and complex scattering geometries \citep{Savic2019,Dotti2022,Marin2023}. Unlike single SMBH systems, which typically show smooth polarization trends dominated by a single disk (see Figure \ref{author_fig1_label}), binary systems can exhibit:
\begin{itemize}
 \item A sudden change in optical/UV flux due to the presence of a gap in the primary SMBH’s accretion disk (the circumbinary disk).
 \item A secondary peak in polarized flux in the FUV, associated with the smaller, hotter accretion disk of the secondary SMBH.
 \item Notable wavelength-dependent polarization position angle variations, indicative of multiple scattering regions with different orientations.
\end{itemize}

These features are not only unique indicators of a binary system, but also serve as direct probes of asymmetric accretion in dynamically evolving environments. The location and intensity of the FUV polarization peak, for example, depends on the mass and accretion properties of the secondary SMBH, providing constraints on how accretion proceeds in non-axisymmetric systems \citep{Milosavljevic2005,Yan2015}. Such measurements are essential to test models of SMBH growth and binary interaction during galaxy evolution.

\begin{figure*}[ht]
\includegraphics[width=\textwidth]{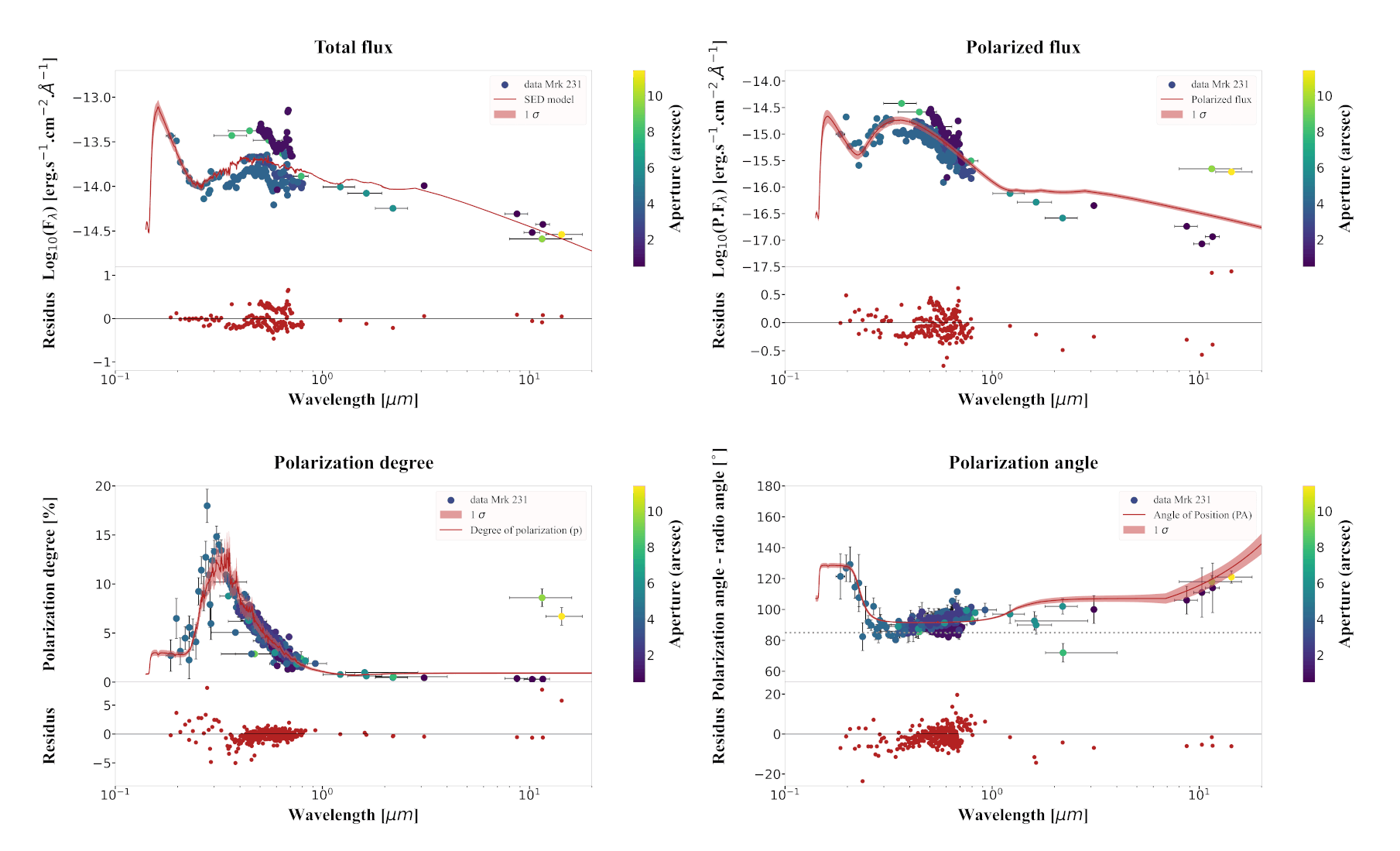}
\caption{\small Mrk 231 total flux (top-left panel), polarized flux (top-right panel), polarization degree (bottom-left panel) and polarization position angle (bottom-right panel) as a function of wavelength. The polarization position angle has been substracted from the observed radio position angle (5$^\circ$, see \citealt{Ulvestad1999}). The solid red line shows a fit of a binary SMBH model, with the residuals between the observational data and the theoretical model plotted below each panel. Figure from \citet{Biedermann2025}.
\label{author_fig2_label}
}
\end{figure*}

Figure \ref{author_fig1_label} compares the expected total and polarized fluxes of an AGN with a binary SMBHs system (red curve) to the same model with a single SMBH (green curve). The binary system exhibits a characteristic UV polarization peak due to the interplay between the primary and secondary SMBH’s accretion disks, a feature that is not apparent in the case of an AGN with a single SMBH. A sharp change in polarization angle occurs at the wavelength where the primary disk's contribution dominates. Those distinct signatures are crucial indicators of a binary SMBH but, if they manifest themselves in the mid- or FUV domains, they are out of reach of modern instruments, that are limited to the near-UV band. This is unfortunate as the most direct observational evidence for a binary SMBH would be a second peak in polarized flux in the FUV, corresponding to the emission from the accretion disk of the secondary black hole (see Figure \ref{author_fig1_label}, top-left panel). As of today, there is only one target for which such specific signatures have been observed : Mrk 231. It is one of the brightest ultraluminous infrared galaxies in the nearby Universe that displays a unique optical-UV spectrum, characterized by a strong attenuation in the near-UV associated with a sudden polarization peak (see Figure \ref{author_fig2_label}).

Figure \ref{author_fig2_label} shows that a binary SMBH model does not only correctly fits the total flux SED but is remarkably well aligned with the observed polarization signatures of Mrk 231 \citep{Biedermann2025}. This is even more impressive considering that the total flux, the polarization degree and the polarization angle are three independent observable, so fitting all three of them at once with the same model provides a strong support for a potential binary SMBH at the heart of Mrk 231. In addition, two bumps are predicted in polarized flux, the first one in the FUV and the other one in the optical. The optical bump is clearly seen in the polarized data but the lack of measurements in the FUV prevents any definite proof of the presence of a secondary, less-massive SMBH. This is a strong prediction that could be potentially examined in the future if a space-based, FUV ($>$ 1000~\AA) polarimeter such as POLLUX is pointed towards this source.

\subsection{Probe the accretion disk physics and red edge}

Accretion disks around supermassive black holes are fundamental to our understanding of AGNs, as they are responsible for the majority of their radiative output. The standard model of optically thick, locally heated disks predicts a specific spectral shape, particularly a blue asymptotic slope in the optical and NIR bands \citep{Pringle1972,Shakura1973,Lynden1974}. However, observational discrepancies have persisted, particularly concerning the spectral shape at longer wavelengths. One major challenge in verifying disk emission in the NIR is the overwhelming contamination from hot dust emission originating from larger, often unresolved scales \citep{Koratkar1999}.

\begin{figure}[ht]
\includegraphics[width=\columnwidth]{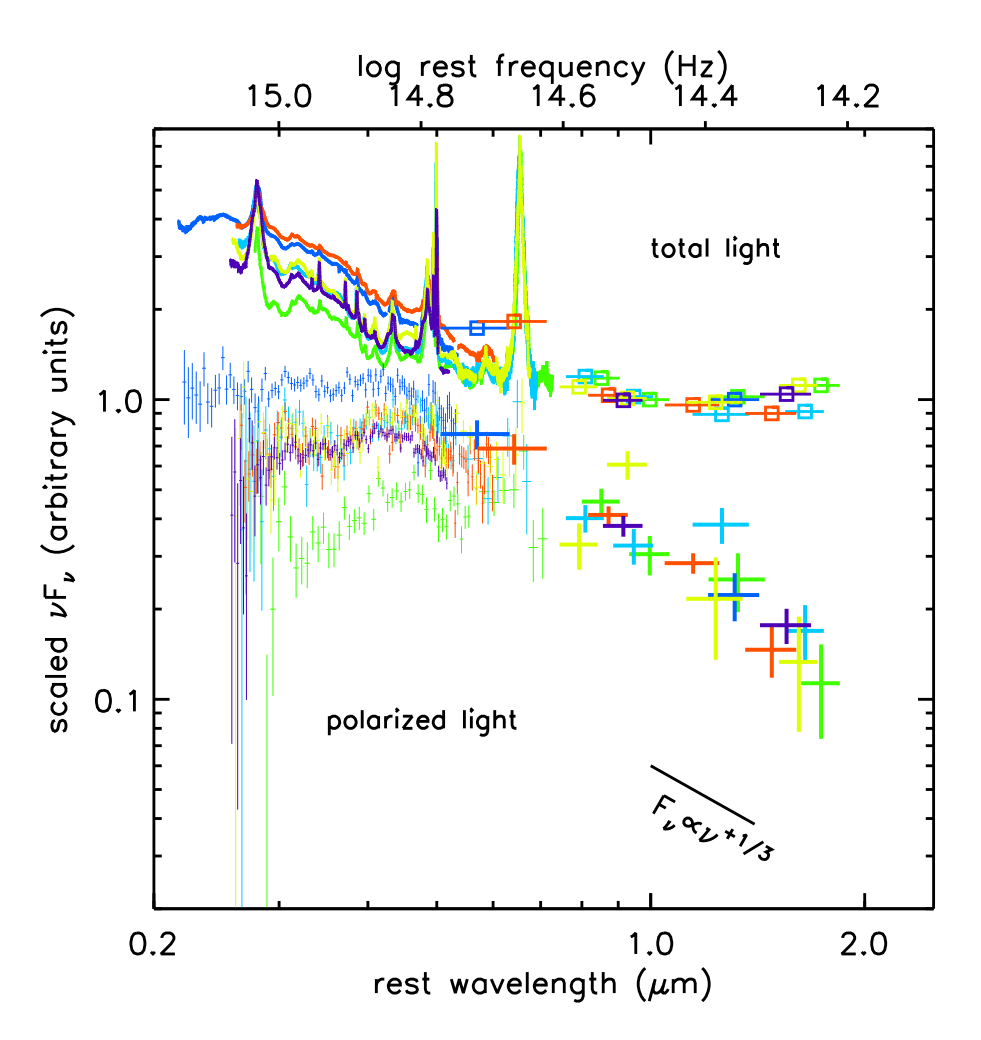}
\caption{\small Overlay of the polarized and total light spectra observed in six different pole-on AGNs. Total light spectra, shown as solide lines in the optical and as squares in the NIR, are normalized at 1~$\mu$m in the rest frame. Polarized light spectra, shown as light points in the optical and as bold points in the optical and NIR both (vertical error bars, 1$\sigma$), are separately normalized, also at 1~$\mu$m. The normalized polarized light spectra are arbitrarily shifted downwards by a factor of three relative to the normalized total light spectra, for clarity. The total light spectra in $\nu$F$\nu$ turns up at around or slightly longward of 1~$\mu$m. In contrast, the polarized-light spectra in $\nu$F$\nu$ all consistently and systematically decrease towards long wavelengths, showing a blue shape of approximately power-law form. Figure taken from \citet{Kishimoto2008}.
\label{author_fig3_label}
}
\end{figure}

This objective aims to characterize the red edge of quasar accretion disks by leveraging observations of polarized light, which enables the isolation of the disk spectrum from the surrounding, little-to-no polarized dust emission. Previous studies \citep{Kishimoto2008,Marin2020} have demonstrated that in several AGNs, the NIR polarized light follows a blue spectral slope consistent with the predictions of the standard geometrically-thin, optically-thick accretion disk model, in contrast to the total-light spectrum, which is dominated by dust emission beyond $\sim$1~$\mu$m. The systematic behavior of the polarized spectra suggests that the intrinsic disk spectrum in the NIR conforms to the expected F$_\nu \propto \nu^{+1/3}$ dependency (see Figure \ref{author_fig3_label}).

By systematically analyzing the NIR polarized light from pole-on AGNs, we could refine our understanding of the structure and physics of accretion disks, providing crucial constraints on their thermal and radiative properties. Key goals include:
\begin{itemize}
 \item measuring the spectral slope of NIR polarized light in a larger sample of quasars to confirm systematic trends.
 \item investigating the radial temperature profile (T $\propto$ r$^{-3/4}$) inferred from the spectral properties of the polarized light.
 \item assessing the stability of the disk at large radii, where gravitational instabilities could lead to spectral deviations and the creation of the broad line emitting region. 
 \item exploring whether the disk exhibits a spectral break at longer wavelengths, potentially indicating a physical truncation or fragmentation and shedding light on the matter supply to the central black hole.
\end{itemize}

This plot illustrates that, under polarized light, the SED of AGN can show the expected F$_\nu \propto \nu^{+1/3}$ dependency, implying that the standard yet controversial picture of the disk being optically thick and locally heated is approximately correct, at least in the outer NIR emitting radii. However, according to theories (see, e.g., \citealt{Lasota2016}), such disk should be gravitationally unstable at large radii. One of the signatures of such break would be a spectrum becoming even bluer at the longest wavelengths\citep{Kishimoto2005}. Examining many AGNs in polarized light, from the FUV to the NIR, would provide critical hints on how and where the disk ends, and how material is being supplied to the nucleus.

\subsection{Characterize the orbital and accretion properties of binary SMBHs}

Binary SMBHs are expected to exhibit periodic variations in their emission due to orbital motion and accretion dynamics. These periodicities arise from several mechanisms, including Doppler boosting, relativistic beaming, and variations in the accretion rate onto each SMBH \citep{Dotti2022}. The detection and characterization of such periodic signatures are crucial for confirming the presence of binary SMBHs and understanding their evolution.

Theoretical models predict that if both SMBHs host accretion disks, their emission should vary periodically on timescales comparable to the orbital period (e.g., $\sim$1 yr for sub-parsec binaries). Optical and UV continuum flux variations can result from the periodic modulation of the accretion rate onto one or both SMBHs. Similarly, the FUV and X-ray emission, originating from the innermost regions of the accretion disks, should display periodic variability due to disk precession and tidal interactions \citep{Liu2023}. Additionally, the emission-line profiles from the broad-line region (BLR) provide critical insights into the kinematics of the system. In a binary SMBH scenario, velocity shifts between different ionization lines are expected due to differential illumination from the individual accretion disks \citep{Gaskell1983,Gaskell2010}. Periodic shifts in the peak or width of emission lines such as Ly$\alpha$ (FUV), H$\beta$ (optical), and He I* (NIR) can serve as a kinematic tracer of orbital motion \citep{Leighly2016}. 

Recent observational studies have provided conflicting results regarding periodic variability in candidate binary SMBH systems. While some long-term monitoring surveys have identified quasi-periodic oscillations in optical photometry and UV spectroscopy, others report a lack of significant variations. For instance, \citet{Kova2020} detected periodic optical variability in Catalina Surveys Data Release 2 using Lomb-Scargle periodograms, suggesting the presence of an orbiting secondary SMBH in Mrk 231. Conversely, \citet{Veilleux2016} reported no significant UV/optical variations over multi-year timescales in the same object, challenging the binary SMBH interpretation.

\begin{figure}[ht]
\includegraphics[width=\columnwidth]{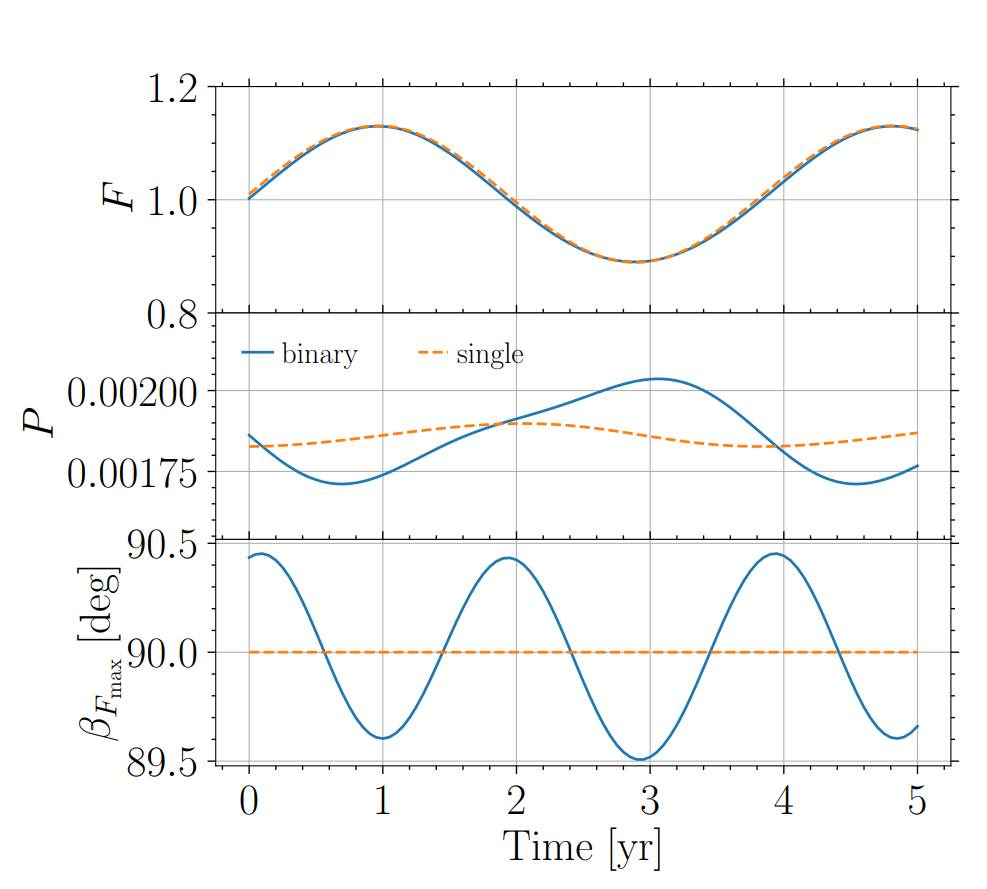}
\caption{\small Light curve simulations of a single (dashed orange line) and a binary (solid blue line) SMBH system. The upper panel is the usual light curve, the middle panel shows the modulation of the polarization fraction and the bottom panel shows the variation in time of the polarization angle. Figure taken from \citet{Dotti2022}.
\label{author_fig4_label}
}
\end{figure}

To resolve these discrepancies, a dedicated monitoring campaign combining photometric, spectroscopic, and polarimetric observations is required. High-cadence (bi-weekly) observations in the FUV are particularly crucial, as they probe the emission from the secondary SMBH, which is expected to be the dominant UV source in many binary models. Key observables are :
\begin{itemize}
 \item Polarization variability of the continuum, as changes in polarization degree and polarization angle over time can reveal the orbital period of the system as well as the inclination of its main constituents (see Figure \ref{author_fig4_label}).
 \item Polarization variability in the lines, since disk precession should induce periodic variations in the polarization angle of broad emission lines, which is directly linked to the binary’s orbital configuration and the gravitational interactions between the two SMBHs. Additionally, because broad emission lines originate from gas moving at high velocities, if the system is a binary, velocity-dependent polarization is also expected, where different velocity components of the line (blue/red wings) show different polarization degrees and angles.
\end{itemize}

Figure \ref{author_fig4_label} shows an example of what kind of temporal modulation one can expect from a system with a binary SMBH in comparison to a regular AGN with a single black hole. A modulation is imparted by the binary motion on the direct light, and, due to the small contribution of scattered light to the total flux, the latter is modulated with the same period, as clearly observable in figure. The middle panel shows the evolution of the polarization fraction $P$, which varies on the same timescale, but which has its maximum close to the minimum of the direct flux (and, therefore, of the total flux), that, being unpolarized, suppresses $P$. The bottom panel shows the evolution of the polarization angle, which oscillates around a central value of 90$^\circ$. Such value is due to (1) the specific orientation of the reference frame chosen (with the x-axis parallel to the semi-major axis of the projected orbital ellipse), and (2) the fact that the the scattered light is more polarized when the scattering angle is closer to 90$^\circ$. One can see that the time-dependent variations of the polarization properties are small. Measuring them requires not only a sensitive and high spectral resolution instrument (for the lines), but also a telescope with a large collective area to maximize the signal-to-noise ratio. By combining high-cadence photometric variability analyses, spectroscopic line-profile monitoring, and polarimetric measurements across a sample of candidate binary SMBHs, it is thus possible to robustly identify binary SMBHs and refine our understanding of their accretion physics.

\section{Physical Parameters}

The ultimate goal of this Signature Science Case is to determine whether binary SMBHs truly exist by measuring the (time-dependent) FUV to NIR polarization, both in continuum and emission lines, together with the total flux of a large sample of AGNs. To do so, the instrument needs to measure the polarization degree and polarization angle of the source, with an uncertainty as small as one hundredth of a percent (for the polarization degree) or one tenth of a degree (for the polarization angle). By doing so, it will be possible to determine the geometric configuration of the central engine: one or two SMBHs, the inclination of their respective accretion disk, the binary’s orbital configuration...

As shown on Figure \ref{author_fig4_label} and in \citet{Savic2019}, the temporal modulation of the polarization, both in degree and angle, is small, less than a fraction of a percent or a degree. This is probably the reason why no periodic modulations have been found in polarimetry so far, the data being hidden behind too large error bars. The cause can be attributed to both dilution of the polarization by the host starlight component and the fact that binary SMBHs can only be probed in AGNs seen close to face-on (i.e. lowly polarized) inclinations, otherwise the observer’s line-of-sight would be blocked by a circumnuclear of dust and gas situated along the AGN equatorial plane and often called “the torus”. High signal-to-noise ratio and state-of-the-art instruments are thus required, coupled to large mirrors (i.e. large effective areas). But the periodic modulations of the polarization is only one face of the coin, as measuring the FUV to NIR continuum and line polarization is also crucial to find wavelength-dependent variations of the polarization signatures. Finally, such measurement, providing for free the total flux, would help to find AGN SEDs with dips in their spectrum, indicating truncation of the circumbinary accretion disk by a second (smaller) black hole and its own disk emitting mainly in the UV. This feature is not so easy to distinguish in total flux, as it can be hidden behind the host starlight SED that can, in some faint AGNs, dominates over the optical (thermal or synchrotron) emission from the compact core \citep{Lofthouse2018}. 

\begin{figure*}[ht]
\includegraphics[width=\textwidth]{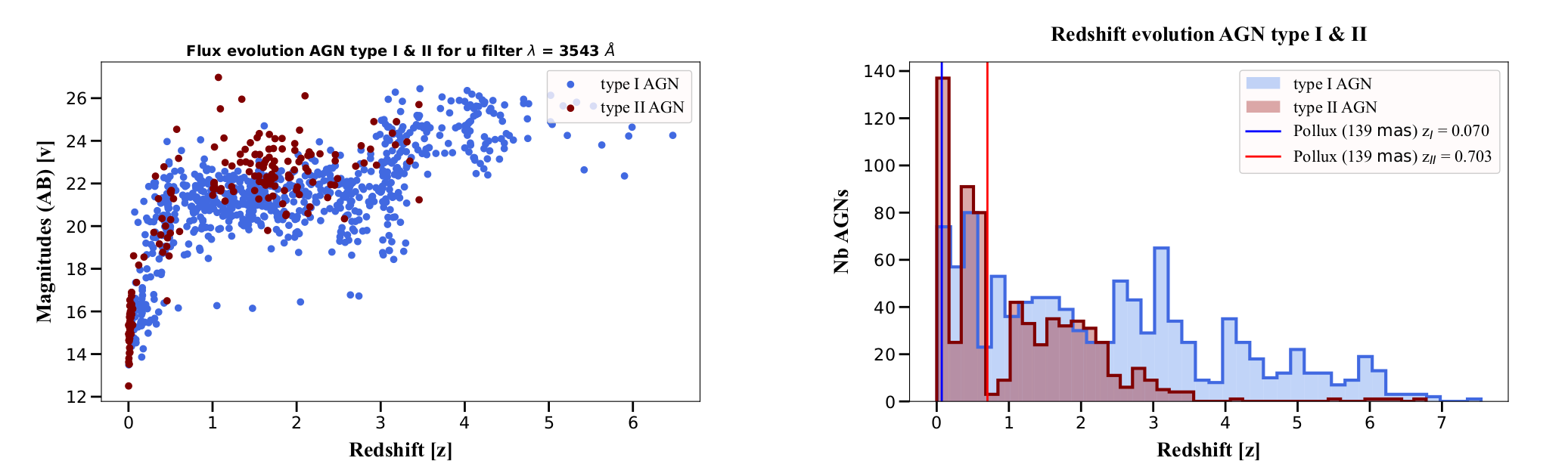}
\caption{\small AGN UV magnitudes as a function of redshift in the SDSS sample (left) and number of AGNs at a given redshift in the same sample (right). Type-1 (face-on) and type-2 (edge-on) AGNs are shown in blue and red respectively. The solid vertical lines delimites the objects that are too big to fit in the pinhole of POLLUX (selected AGNs must then be on the right of the said curve).
\label{author_fig5_label}
}
\end{figure*}

So far, only one object (Mrk 231) has been identified with a clear dip in its SED and another candidate (SDSS J025214.67-002813.7, \citealt{Liao2021}) is yet to be investigated, both in total flux in the UV and in polarization in general (no recorded polarimetric measurement at all). A large sample of type-1 AGNs (see Figure \ref{author_fig5_label}), situated at various redhsifts must be examined to look for potential signatures of a binary SMBH system in their total and polarized flux spectra (isolated peaks of polarization, rotations of the polarization angle before and after the dip...), before initiating intensive (bi-weekly) monitoring in polarimetry. Previous catalogs of polarimetric campaigns of local AGNs are all limited to near-UV wavelengths at the shortest wavelengths and their polarized spectra (if any) have too large error bars to detected anything. The few AGNs that benefited for past FUV polarimetric campaigns (thanks to the WUPPE and the HST/FOS) do not show the tell-tale signatures that are looked for in this proceeding.

Identifying one or more new binary SMBH candidates based on their total and polarized flux spectra is critical, since Mrk 231 is the only known object so far. To achieve a new binary SMBH detection at 95\% confidence level, accounting for an occurrence rate between 0.1 and 1\% \citep{Graham2015,Wang2017}, between 300 and 3000 AGNs have to be observed randomly, unless suspicions of the presence of a binary SMBH (from other observational methods) are provided upstream for certain objects. Collecting all SMBH binary candidates from the literature beforehand and focusing on these objects during a first observation campaign would drastically reduce the quantity of AGNs to be observed and would easily allow imposters to be rejected.

In all cases, reducing the wavelength coverage would be strongly detrimental to the science, since either the secondary peak of UV emission would be lost, or the red edge of the disk would be left unexplored, harming the interpretation and characterization of any discovered binary system. Finally, uncertainties on the integrated polarization (in a waveband without contamination by strong unpolarized emission lines) larger than a hundredth of a percent (in polarization degree) would likely miss the polarization modulation. Errors of the order of one tenth of a percent are already easily achievable with current ground telescopes and an order of magnitude increase is very reasonable for a mission such as the Habitable Worlds Observatory.

\begin{table*}[!ht]
\caption{Quantifying the objectives.}
\smallskip
\begin{center}
{\small
\begin{tabular}{lcccc}  
\tableline
\noalign{\smallskip}
Number of AGNs with... & State of the Art & Incremental Progress & Substantial Progress & Major Progress\\
~ & ~ & (Enhancing) & (Enabling) & (Breakthrough)\\
\noalign{\smallskip}
\tableline
\noalign{\smallskip}
known SED dip & 1 & 2 & 5 & 10 \\
remarkable polarization peaks and $\lambda$-dependance & 1 & 2 & 5 & 10 \\
known polarized periodicity & 0 & 1 & $>$1 & $>$1 \\
polarized FUV-to-NIR spectrum & 12 & 20 & 50 & 100 \\
\tableline\
\end{tabular}
}
\end{center}
\end{table*}

\section{Description of Observations}

To get the measurement done, it is necessary to observe at least 300 AGNs at a redshift large enough so that the entire AGN core (including the polar winds) is inscribed in the aperture of the instrument. The SDSS catalog contains more than enough potential targets, as shown on Figure \ref{author_fig5_label}, with UV magnitudes lower than 22. A signal-to-noise ratio of 500 is necessary, as the usual levels of polarization of the continuum is less than a percent in most type-1 AGNs. Modest spectral resolution is enough (R $>$ 20 000) as the most important lines are broad ($>$ 1000 km~s$^{-1}$). 

Observations should be carried at once from the FUV (1000~\AA) to the NIR (2~$\mu$m) in spectropolarimetry, so that the entire spectrum is observed without time delay and under the same conditions. Former binary SMBHs candidates and bright type-1 (pole-on) sources will be priorities since polarimetry is photon-intensive. The sample of observed AGN will be analyzed gradually, so that whenever a serious binary SMBH candidate is found, bi-weekly (or longer) monitoring onsets. This cadence, of course, will depend on the estimated orbital configuration and periodocity of the system, inferred from the polarized SED. Polarized spectra obtained for rejected binary candidates will not be wasted, since they will be examined in their red part to determine if outer edge disk truncation or fragmentation is observed. Their spectral slope and their radial temperature profile inferred from the spectral properties of the polarized light will be used to put constraints on accretion disk theories (mass transfer functions, disk formation, disk instabilities, disk atmosphere...).

The use of a space-based observatory with a large collecting area is crucial, since the FUV band is not accessible from Earth and requires large mirrors to collect enough photons for statistically significant measurements. The high resolution, broadband ESA/CNES spectropolarimeter POLLUX is, currently, the only proposed instrument that could observe the FUV-to-NIR polarization of extragalactic sources.

\begin{table*}[!ht]
\caption{Instrument design specifications.}
\smallskip
\begin{center}
{\small
\begin{tabular}{lcccc}  
\tableline
\noalign{\smallskip}
Observation Requirement & State of the Art & Incremental Progress & Substantial Progress & Major Progress\\
~ & ~ & (Enhancing) & (Enabling) & (Breakthrough)\\
\noalign{\smallskip}
\tableline
\noalign{\smallskip}
Error on polarization measurement & 0.1\% (per spectral bin) & 0.08\% & 0.05\% & 0.01\% \\
Wavelength Range & 1150~\AA\, -- 0.8~$\mu$m) & 1000~\AA\, -- 1~$\mu$m) &  1000~\AA\, -- 1.5~$\mu$m) & 1000~\AA\, -- 2~$\mu$m) \\
Resolving power & 20\,000 & $>$30\,000 & $>$40\,000 & $>$50\,000 \\
\tableline\
\end{tabular}
}
\end{center}
\end{table*}

\bibliography{author.bib}

\end{document}